\documentclass[twocolumn,showpacs,preprintnumbers,amsmath,amssymb,prb]{revtex4}
\usepackage{graphicx}% Include figure files
\usepackage{dcolumn}% Align table columns on decimal point
\usepackage{bm}% bold math

\begin{document}

%\preprint{APS/123-QED}

\title{Parallel quantized charge pumping}

\author{S.~J.~Wright$^{1,3}$, M.~D.~Blumenthal$^{1}$, M.~Pepper$^{2,1}$, D.~Anderson$^{1}$, G.~A.~C.~Jones$^{1}$, C.~A.~Nicoll$^{1}$, D.~A.~Ritchie$^{1}$\\
}

%\affiliation{Affiliations}
\affiliation{
${}^1$\footnotesize Cavendish Laboratory, University of Cambridge, J. J. Thomson Avenue, Cambridge CB3 0HE, UK.
${}^2$\footnotesize Department of Electronic \& Electrical Engineering, University College London, Torrington Place, London WC1E 7JE, UK.
${}^3$\footnotesize Toshiba Research Europe Ltd, Cambridge Research Laboratory, 208 Science Park, Milton Road, Cambridge CB4 0WE, UK.
}

\date{\today}% It is always \today, today,

\begin{abstract}

Two quantized charge pumps are operated in parallel. The total current generated is shown to be far more accurate than the current produced with just one pump operating at a higher frequency. With the application of a perpendicular magnetic field the accuracy of quantization is shown to be $<\,$20 ppm for a current of $108.9\,$pA. The scheme for parallel pumping presented in this work has applications in quantum information processing, the generation of single photons in pairs and bunches, neural networking and the development of a quantum standard for electrical current. All these applications will benefit greatly from the increase in output current without the characteristic decrease in accuracy as a result of high-frequency operation.

\end{abstract}

\pacs{Valid PACS appear here}% PACS, the Physics and Astronomy
\maketitle

Quantum metrology aims to redefine the Ampere in terms of true invariants of nature. High-accuracy quantized transport of electrons in semiconductor heterostructures is one potential solution that has gained wide attention in recent years~\cite{geerligs1990,andregg1990,kouwenhoven1PBI,pothier1PBI,piquemal04}. The total current produced in such a device is given by $I_{\mathrm{tot}}=nef$, where $f$ is the frequency of operation, $n$ is the number of electrons transported per cycle and $e$ is the elementary charge. For a practical standard, a current of a nanoampere or above with an accuracy of greater than one part per million (ppm) is required. In a multiple tunnel junction device~\cite{keller}, an error of 15 parts in 10$^{9}$ was reported but the frequency of operation was limited to $f\leq20\,$MHz. The level of current generated was shown to be sufficient for a capacitance standard~\cite{keller2} but too small to  be a practical current standard. Surface-acoustic-wave (SAW) driven electron pumps~\cite{Shilton1996} produce higher currents, but their accuracy has not yet been demonstrated to better than one part in $10^{4}$. Square-wave driven charge coupled devices (CCDs)~\cite{fujiwara04} fabricated in silicon operate at frequencies of up to $f=100\,$MHz with a resulting current of $\sim16\,$pA. Higher currents would require better transmission of the sharp pulses to the gates. More recently, J.~P.~Pekola \emph{et al} presented a scheme for generating quantised current in a hybrid single-electron transistor comprising two normal-metal leads and a superconducting central island, with a potential accuracy of 10$\,ppb$~\cite{pekola08_prl}. Preliminary experiments showed quantisation limited to one part in $10^{3}$ due to subgap leakages~\cite{pekola08}.

The first realization of gigahertz electron pumping without the use of SAWs~\cite{Blumenthal2007} showed degradation in the accuracy of the pumped current with an increase in the operating frequency. At $f=547\,$MHz the accuracy was approximately one part in $10^{4}$. This approach was subsequently simplified by Kaestner \emph{et al} in the form of a single parameter pump~\cite{kaestner:153301}. The inclusion of a perpendicular magnetic field~\cite{wright08,kaestnerB} greatly enhanced the confinement of the electrons, leading to a reduction in the pumping error mechanisms. Parallelization of multiple pumps has been suggested as a possible solution to achieve the required accuracy and current for a standard~\cite{Blumenthal_parallel}. A system of multiple pumps in parallel would also have applications in neural networking where the output of one pump is controlled by the throughput of another. If the parallel pumps are made to deliver single electrons simultaneously from an n-type region to a p-type region then pairs and bunches of photons could be generated at will. Our scheme can be applied to produce an arbitrarily high current at high accuracy and will lead to the development of a quantum standard for current, with benefits in many areas including quantum information processing where the manipulation of individual charges is required.

%We have previously developed an electron pump~\cite{Blumenthal2007} which has been shown to generate quantized current at frequencies of up to $f=3.4\,$GHz. The accuracy of the pumped current was shown to degrade with frequency and at $f=547\,$MHz the accuracy was $\sim$1 part in $10^{4}$. This approach was subsequently simplified by Kaestner \emph{et al} in the form of a single parameter pump~\cite{kaestner:153301}. Under an applied perpendicular magnetic field the degree of quantization is enhanced~\cite{wright08}. Parallelization of multiple pumps has been suggested as a possible solution to achieve the required accuracy and current~\cite{Blumenthal_parallel}. A system of multiple pumps in parallel would also have applications in neural networking where the output of one pump is controlled by the throughput of another. There is also the possibility of producing parallel single photons by transporting the electrons across p-n junctions. Development of an accurate current standard will have benefits in the field of quantum information processing, where the manipulation of individual charges is required.

%Here we present an experimental investigation of a system of two electron pumps operating in parallel. The combined error of the two pumps is shown to be far less than the error arising when one pump is operated at a higher frequency. 

\begin{figure*}[ht]
\begin{minipage}[b]{0.3\linewidth}
%\centering
\includegraphics[scale=0.4]{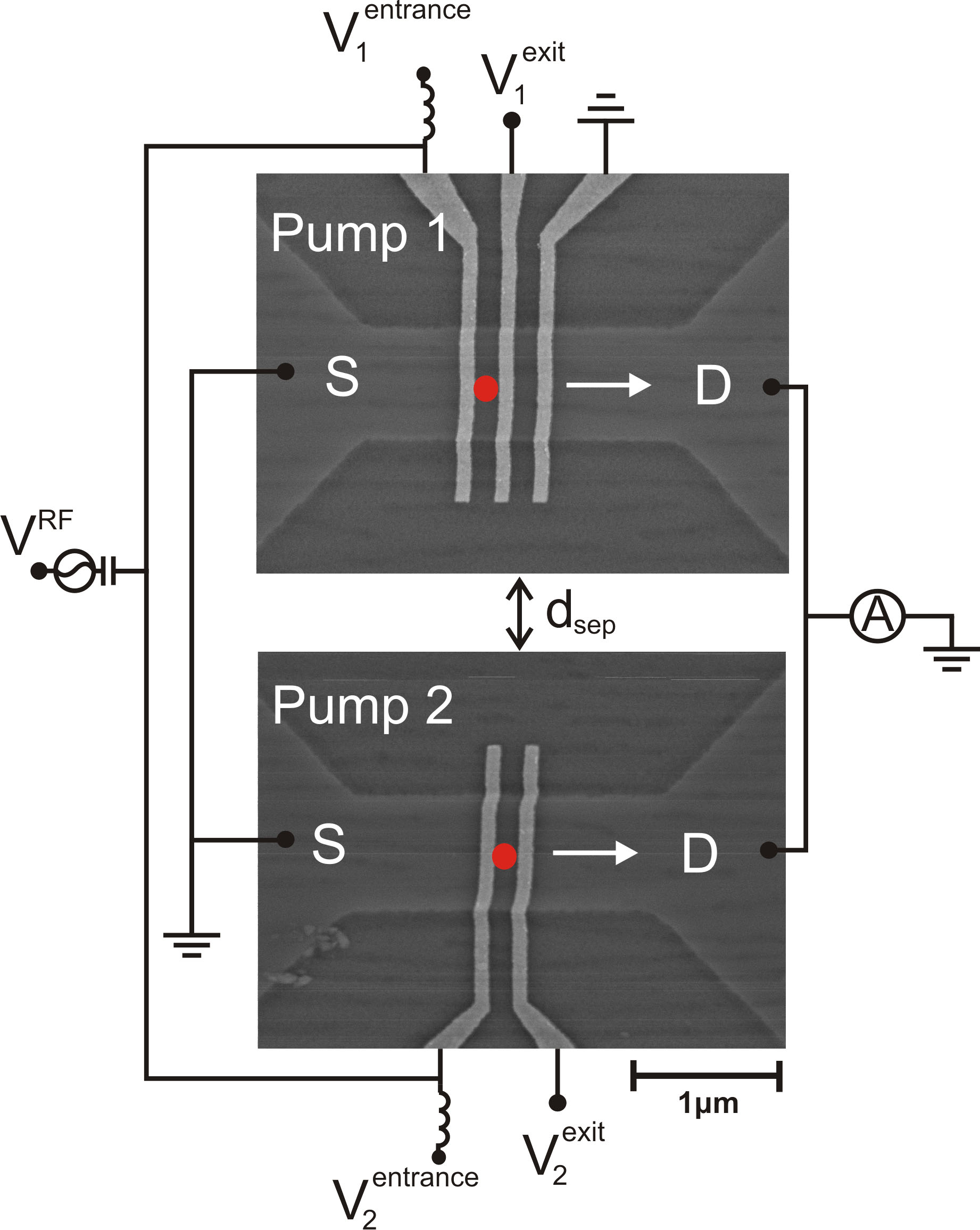}
\caption{(Color online) Scanning electron microscopy image of the device and schematic of electrical connections. The source and drain reservoirs are linked outside the dilution refrigerator. The same sinusoidal RF voltage signal is used to drive both pumps. Static offset voltages $V^{\mathrm{entrance}}_{1}$ and $V^{\mathrm{entrance}}_{2}$ are added to this RF signal.}
\label{fig:device}
\end{minipage}
\hspace{0.5cm}
\begin{minipage}[b]{0.5\linewidth}
%\centering
\includegraphics[scale=0.7]{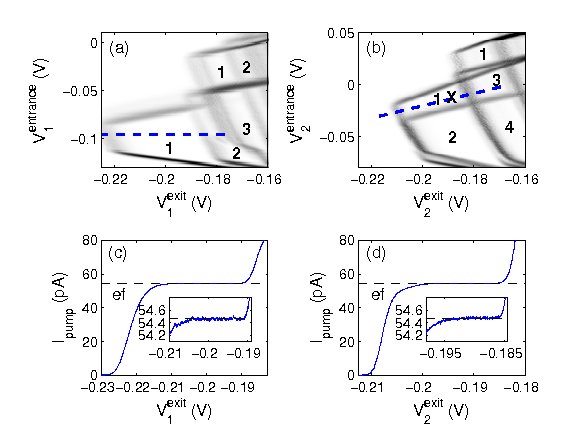}
\caption{(Color online) Pumped current for the range of gate voltages at which a quantized number $n$ of electrons were transported per cycle of the RF signal. (a) and (b) show the numerical derivative of the pumped current for pump 1 and pump 2 respectively in an applied field of $B=5\,$T. Annotations are explained in the text. (c) and (d) show linescans across the first plateau. The insets are magnifications of the plateau regions.}
\label{fig:2D}
\end{minipage}
\end{figure*}

%\begin{figure}[h]
%\includegraphics[width=0.45\textwidth]{figures/device_large.eps}
%\caption{\label{fig:device} {\bf Scanning electron microscopy image of the device and schematic of electrical connections}. The source and drain reservoirs in each case are linked outside the dilution refridgerator. The same sinusoidal RF voltage signal is used to drive both pumps. Static offset voltages $V^{1}_{\mathrm{entrance}}$ and $V^{2}_{\mathrm{entrance}}$ are added to this RF signal.}
%\end{figure}

In this Brief Report we present an experimental investigation of a system of two electron pumps operating in parallel. This was realized by simultaneously forming two few-electron dynamic QD potentials in a GaAs/AlGaAs high electron mobility transistor (HEMT) heterostructure, where a two-dimensional electron gas (2DEG) resides $90\,$nm below the surface. The 2DEG had a mobility of $101.7\,$m$^2/$Vs and an electron density of $1.5\times10^{15}\,$m$^{-2}$ at $1.5\,$K. We created two separate mesas on the same chip, and for each mesa we defined a quasi-1D channel by shallow wet chemical etching. The QDs were created by applying voltages to surface finger gates lying perpendicular to the channels. A scanning electron microscope (SEM) image of the pumps is presented in Fig.~\ref{fig:device}. The red dots indicate the positions at which the dynamic QDs were periodically formed. The etched quantum wires were of lithographic width $w=700\,$nm and length $l=2\,\mu$m, each end opening up to large areas of 2DEG where Ohmic contacts were made. The surface finger gates were formed by evaporation of Ti($10\,$nm)/Au($20\,$nm) using electron beam lithography and standard liftoff techniques; they were of width $100\,$nm and pitch $250\,$nm.

The scheme for parallel pumping suggested in ref.~\cite{Blumenthal_parallel} requires the dynamic QDs to be in close proximity so that each QD can be controlled by common gates. From studies of these systems we found that having the two QDs in such close proximity leads to a large amount of crosstalk of the RF signal interfering with the nearby pump, as well as intra-dot capacitive coupling. These effects make it very hard to tune both pumps simultaneously. The suggestion of using common gates to control many pumps in parallel is also disadvantageous as the gate voltages necessary to exhibit quantized current are shown in this work to be quite different for each pump. We achieved parallel pumping through separating each pump by a distance of $d_{\mathrm{sep}}=160\,\mu$m and using separately controllable gates to define each QD. The design allows total independence in tuning the individual pumps and we no longer suffer from the effects of crosstalk and capacitive coupling.

To induce parallel pumping, static voltages $V^{\mathrm{entrance}}_{1}$, $V^{\mathrm{exit}}_{1}$, $V^{\mathrm{entrance}}_{2}$ and $V^{\mathrm{exit}}_{2}$ were applied, as indicated in Fig.~\ref{fig:device}. The oscillating voltage signal $V^{\mathrm{RF}}$ was split at room temperature by a splitter, {\bf Sp}, and used to drive both pumps. This RF signal was added to the entrance gate of each pump by means of bias tees, as shown. When tuned correctly, a quantized number of electrons can be captured from each source lead by the QDs and deposited into the respective drain in every cycle of the RF signal. These currents were added and measured using a Keithley 6430 electrometer.

Figures~\ref{fig:2D}(a) and (b) show the numerical derivative of the pumped current as a function of each gate voltage for pump 1 and pump 2 respectively. All measurements were made in a dilution refrigerator at a temperature of $\sim40\,$mK. The RF signal was set to a frequency of $f=340\,$MHz. A perpendicular magnetic field of $B=5\,$T was applied. The plateaus that formed are the signature of quantized current. Numbers in the plateaus correspond to the number $n$ of electrons transported per cycle of the RF signal for those regions of gate voltage. The dashed blue lines indicate the direction of linescans across the first plateau for each pump. These linescans are presented in Figs.~\ref{fig:2D}(c) and (d). The insets are magnified plots of the plateau regions. At this $B$ field, the plateaus corresponding to one electron pumped per cycle had a gradient which was smaller than its error. It can therefore be said that the plateaus were flat to within the resolution of the measurement system. With the electrometer set to the $100\,$pA range this resolution was found to have an upper limit of $\sim0.7\,$fA, giving a current which is invariant in gate voltage to better than $12\,$ppm at $340\,$MHz~\cite{absolute}. Black dashed lines indicate the expected current $I=ef$ using the value of $e$ recommended by the Committee on Data for Science and Technology (CODATA).

\begin{figure}
\includegraphics[width=0.45\textwidth]{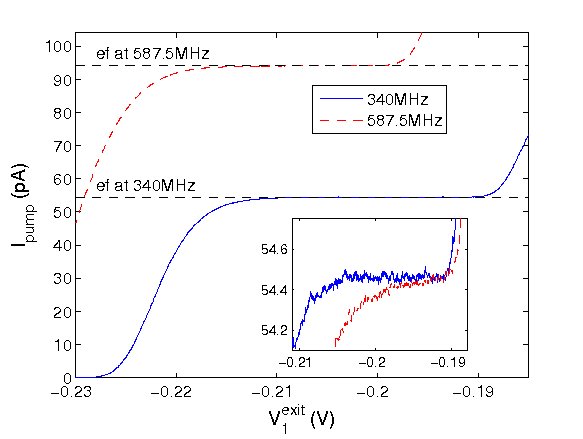}
\caption{\label{fig:freq}(Color online) Pumped current for the first plateau at 340$\,$MHz and 587.5$\,$MHz. The inset shows the plateaus magnified and normalized in gate voltage. The plateau at 587.5$\,$MHz has been scaled to $ef$ at 340$\,$MHz for comparison.}
\end{figure}

For frequencies above 340$\,$MHz the plateaus appeared noticeably sloped. Figure~\ref{fig:freq} shows the first plateau for pump 1 at a frequency of 340$\,$MHz and 587.5$\,$MHz. The inset is a magnified plot of each plateau, normalized in gate voltage and scaled to $ef$ at 340$\,$MHz for comparison. This figure illustrates the motivation for this work. The plateau at 587.5$\,$MHz is clearly sloped, indicating a marked increase in the error associated with pumping at such a frequency. It is also clear that the plateau at 587.5$\,$MHz is shorter than the plateau at 340$\,$MHz, indicating a reduction in the robustness. 

Due to cable resonances we found that only narrow-band frequency ranges showed sufficient transmission of the RF signal to the gates in order to induce pumping. All such frequencies above 340$\,$MHz showed noticeably sloped $ef$ plateaus. At higher frequencies the potential for nonadiabatic excitations of the confined electrons is increased which is likely to have an adverse effect on the quantization, leading to an increased probability for back tunneling of the captured electrons to the source reservoir. When more than one pump is operated in parallel at lower frequencies these nonadiabatic effects are suppressed and a higher accuracy results.

\begin{figure}
\includegraphics[width=0.45\textwidth]{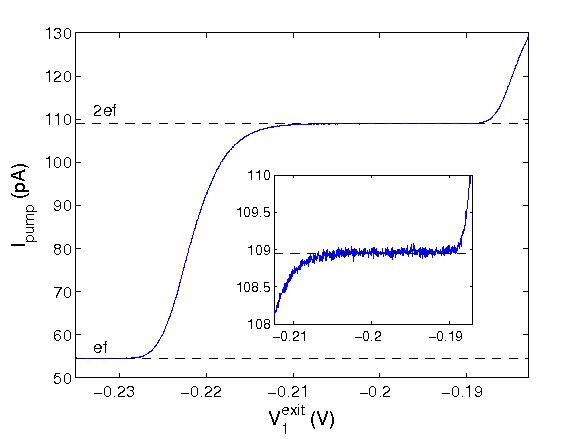}
\caption{\label{fig:parallel}(Color online) Pumped current when both pumps are operating in parallel. The total current produced is $2ef$. The inset is a magnified plot of the plateau region. }
\end{figure}

In order to tune for parallel pumping we set gate voltages $V^{\mathrm{entrance}}_{2}$ and $V^{\mathrm{exit}}_{2}$ such that pump 2 exhibited the first plateau. These voltages are shown by the cross in Fig.~\ref{fig:2D}(b). We then repeated the sweep of the exit gate for pump 1 shown in Fig.~\ref{fig:2D}(c). The resulting pumped current is presented in Fig.~\ref{fig:parallel}. When both pumps are tuned to the first plateau a total current of $2ef$ (108.95pA) is produced. With the electrometer set to the 1$\,$nA range the resolution of the measurement system was $\sim2\,$fA, so that the pumped current was invariant in gate voltage to better than $20\,$ppm. Janssen and Hartland~\cite{janssen} have demonstrated that the flatness of the quantized current plateau is empirically related to the accuracy of quantization. Through operating two pumps in parallel we have therefore achieved the highest degree of quantization reported to date in this current range. 

It is observed from the inset of Fig.~\ref{fig:parallel} that the absolute value of the plateau is slightly higher than expected. We intend to measure the plateau value using a calibrated current source to quantify and explain any real deviation from the expected value indicated by the dashed black line. This will be the subject of future work. Figure~\ref{fig:parallel} demonstrates conclusive proof that two pumps can be operated in parallel without loss of quantization.

In conclusion, we have demonstrated parallel pumping using gated GaAs/AlGaAs dynamic QDs. Our scheme involves spatially separating the pumps and controlling them wth separate gates, overcoming technical issues pertaining to a previously outlined scheme~\cite{Blumenthal_parallel} where it was suggested that multiple pumps could be operated in close proximity with common gates. Application of a perpendicular magnetic field greatly improved the plateau flatness, hence accuracy of quantization of the pumped current, to better than $20\,$ppm. Using the scheme for parallelization outlined in this work, many electron pumps could be operated in parallel without loss of quantization, with the total output current limited only by resources. This work demonstrates the operation of an electron pump system with all the building blocks necessary for realising an electrical standard for current based on true invariants of nature.

%In conclusion, we have demonstrated parallel pumping using gated GaAs/AlGaAs dynamic QDs. The system for parallel pumping proposed in ref.~\cite{Blumenthal_parallel} is shown to be unsuitable. In this scheme the QDs are brought close to each other so that common gates may be used to drive both pumps. This approach is unlikely to produce a high yield of working devices because the tuning of the gates tends to be quite different for each pump. The pumps we present in this work needed quite different gate voltages $V^{1}_{\mathrm{entrance}}$ and $V^{2}_{\mathrm{entrance}}$ to produce quantised current. A common gate limits the parameter space for finding pumping in both QDs. Our scheme allows for more flexibility in the tuning of the devices and can be adapted to accommodate an arbitrary number of pumps operating in parallel, paving the way for the realisation of an electrical standard for current based on true invariants of nature.

We gratefully acknowledge fruitful discussions with T.~J.~B.~M.~Janssen and S.~Giblin of the National Physical Laboratory. This work was supported by the EPSRC. SJW acknowledges additional support from Toshiba Research Europe Ltd as part of an industrial CASE award.

%CN acknowledges support from the EPSRC QIP IRC (GR/S82176/01).


\begin{thebibliography}{99}
\bibitem{absolute} Without a calibrated current meter we cannot make any claims about the absolute value of the plateaus. This will be the subject of future work.
\end{thebibliography}
\end{document}